\input{aipcheck}

\documentclass[
    ,final            
  ]
  {aipproc}

\layoutstyle{6x9}

     \newcommand{\la}{\,\rlap{\raise 0.5ex\hbox{$<$}}{\lower 1.0ex\hbox{$\sim$}}\,}
     \newcommand{\ga}{\,\rlap{\raise 0.5ex\hbox{$>$}}{\lower 1.0ex\hbox{$\sim$}}\,}
\newcommand\fd{\hbox{$.\!\!^{\reset@font\romn d}$}}
\newcommand\fh{\hbox{$.\!\!^{\reset@font\romn h}$}}
\newcommand\fm{\hbox{$.\!\!^{\reset@font\romn m}$}}
\newcommand\fs{\hbox{$.\!\!^{\reset@font\romn s}$}}

\newcommand\fp{\hbox{$.\!\!^{\reset@font\reset@font\scriptscriptstyle\romn p}$}}

\def\ktbb{kT_{\rm bb}}

\def\kte{kT_{\rm e}}
\def\scw{Schwarzschild}

\def \comptt {\textsc{comptt}}
\def \compmag {\textsc{compmag}}

\def\b0{\beta_{\rm 0}}

\def \sw {{\em Swift}}
\def \apj {ApJ}

\def \aap {A\&A}

\def \mnras {MNRAS}

\begin{document}

\title{Supergiant Fast X-ray Transients with \emph{Swift}: 
spectroscopic and temporal properties}


\author{P.~Romano}{ address={Istituto di Astrofisica Spaziale e Fisica Cosmica,        Via U.\ La Malfa 153, I-90146 Palermo, Italy}}
\author{V.~Mangano}{ address={Istituto di Astrofisica Spaziale e Fisica Cosmica,      Via U.\ La Malfa 153, I-90146 Palermo, Italy}}
\author{L.~Ducci}{  address={Institut f\"ur Astronomie und Astrophysik,         Universit\"at T\"ubingen, Sand 1, D-72076 T\"ubingen, Germany }}
\author{P.~Esposito}{  address={Istituto di Astrofisica Spaziale e Fisica Cosmica,         Via E.\ Bassini 15,   I-20133 Milano,  Italy}}
\author{R.~Farinelli}{  address={Dipartimento di Fisica, Universit\`a di Ferrara, via Saragat 1, 44122, Ferrara, Italy}}
\author{C.~Ceccobello}{  address={Dipartimento di Fisica, Universit\`a di Ferrara, via Saragat 1, 44122, Ferrara, Italy}}
\author{S.~Vercellone}{address={Istituto di Astrofisica Spaziale e Fisica Cosmica,       Via U.\ La Malfa 153, I-90146 Palermo, Italy}}
\author{D.N.~Burrows}{ address={Department of Astronomy \& Astrophysics, Pennsylvania State         University, University Park, PA 16802, USA}}
\author{J.A.~Kennea}{ address={Department of Astronomy \& Astrophysics, Pennsylvania State         University, University Park, PA 16802, USA}}
\author{H.A.~Krimm,}{ address={NASA/Goddard Space Flight Center, Greenbelt, MD 20771, USA}}
\author{N.~Gehrels}{ address={NASA/Goddard Space Flight Center, Greenbelt, MD 20771, USA}}

\begin{abstract}
 Supergiant fast X-ray transients (SFXTs) are a class of high-mass X-ray binaries  
with possible counterparts in the high energy gamma rays. 
The \emph{Swift} SFXT Project\footnote{SFXT Project Page: \texttt{http://www.ifc.inaf.it/sfxt/ }  }
has conducted a systematic investigation of the properties 
of SFTXs on timescales ranging from minutes to years and in several intensity 
states (from bright flares, to intermediate intensity states, and down 
to almost quiescence). We also performed broad-band spectroscopy of outbursts, 
and intensity-selected spectroscopy outside of outbursts.  
We demonstrated that while the brightest phase of the outburst only 
lasts a few hours, further activity is observed at lower fluxes for a remarkably 
longer time, up to weeks. Furthermore, we assessed the fraction of the time these 
sources spend in each phase, and their duty cycle of inactivity. 
We present the most recent results from our investigation. The spectroscopic 
and, most importantly, timing properties of SFXTs we have uncovered with \emph{Swift}  
will serve as a guide in search for the high energy emission from these enigmatic 
objects.
\end{abstract}

\maketitle


             \section{Supergiant fast X-ray transients}

Supergiant fast X-ray transients (SFXT) are detected as fast hard X-ray transients
whose hour-long outbursts reach $10^{36}$--$10^{37}$ erg s$^{-1}$ and whose spectra 
resemble those of accreting neutron stars (NS). Indeed, some SFXT are shown to have pulsations
ranging from a few seconds to $\sim 1000$\,s. This flaring gives SFXTs quite a large dynamical range,
which ranges between 3 and 5 orders of magnitude in flux. Optical identification of the counterparts
has led to classify SFXTs as a peculiar class of high mass X--ray binaries (HMXB), associated to a 
blue (OB) supergiant companion. We generally distinguish between 
confirmed and candidate SFXTs based on the availability of an optical classification of the companion,
and we now have 10 confirmed and as many candidate SFXTs. 
The details of the physical origin of the outbursts are still unclear, but the ingredients that need
to be taken into account include the properties of 
the wind from the supergiant companion \cite[][]{zand2005,Sidoli2007} and the 
presence of a centrifugal or magnetic barrier \cite[][]{Grebenev2007,Bozzo2008}.

             \section{The SFXT Project: recent results}

Since 2007 the SFXT Project has performed an investigation of the properties of SFXTs 
with \emph{Swift},  
whose fast-slewing capability and broad-band energy coverage, combined with its 
flexible observing scheduling, make it an ideal facility to study both the bright 
outburst and the out-of-outburst behavior.  
Our strategy pairs monitoring programs with outburst follow-ups 
(see, e.g.\  \cite{Romano2011:TEXAS2010}) and has allowed us to 
catch several outbursts from most confirmed and several candidate SFXTs  
(see, e.g.\ \cite{Romano2011:sfxts_paperVII} and references therein), 
to monitor them during their evolution, 
and to study the long term properties 
of this class of objects with a highly sensitive soft X--ray telescope. 

\begin{figure}
\hspace{-0.7truecm}
   \includegraphics[angle=270,width=.7\textheight]{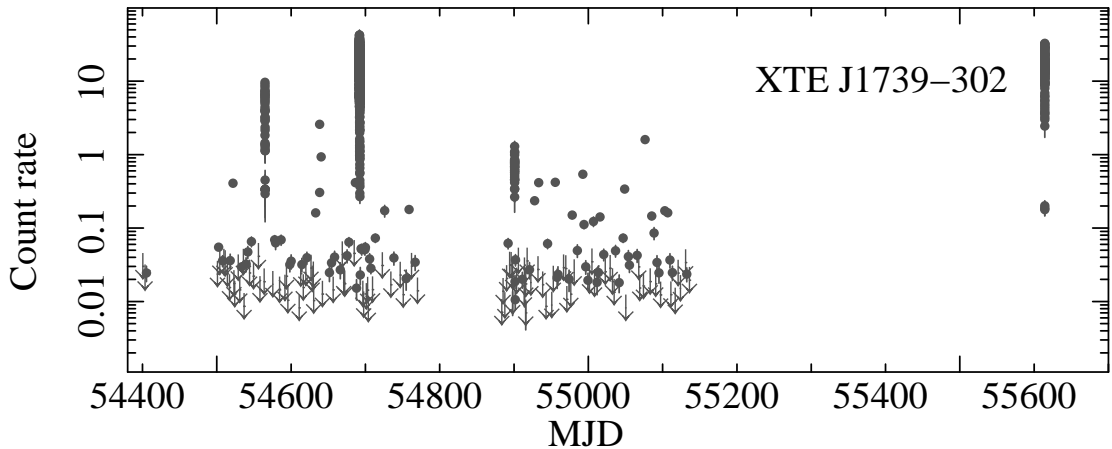}
\caption{\sw/XRT (0.2--10\,keV) long-term light curves of XTE~J1739$-$302 and subsequent outbursts.   
		The downward-pointing arrows are 3$\sigma$ upper limits. 
                Adapted from \cite[][]{Farinelli2012:sfxts_paperVIII}. 
}
  \label{fig1}
\end{figure}

\begin{figure}
 \hspace{-0.7truecm}
  \includegraphics[angle=270,width=.7\textheight]{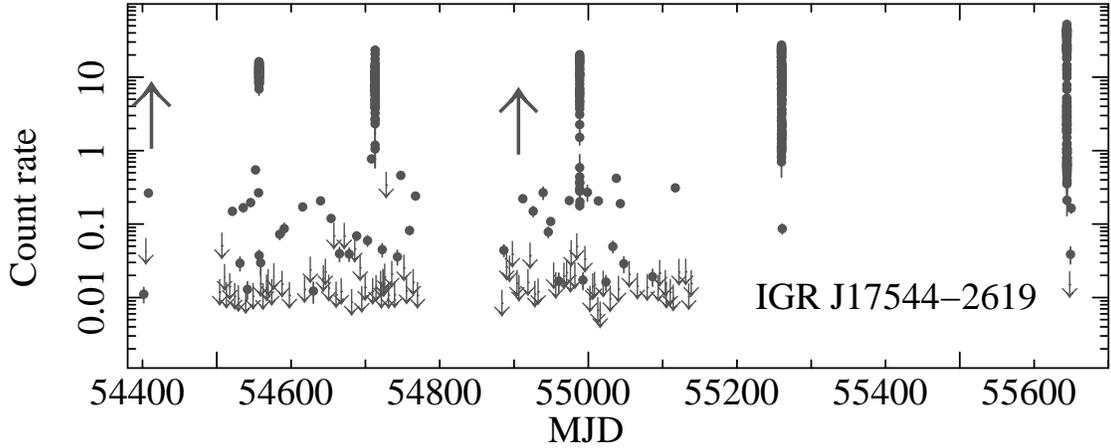}
\caption{\sw/XRT (0.2--10\,keV) long-term light curves of  IGR~J17544$-$2619
and subsequent outbursts.   
		The downward-pointing arrows are 3$\sigma$ upper limits. The upward-pointing arrows 
                mark flares that triggered the BAT Transient Monitor on MJD 54414 and 54906.  
                Adapted from \cite[][]{Farinelli2012:sfxts_paperVIII}. 
}
  \label{fig2}
\end{figure}

Our long-term monitoring of 4 SFXTs (IGR~J16479$-$4514, XTE~J1739$-$302, 
IGR~J17544$-$2619, AX~J1841.0$-$0536) 
has allowed us to determine that they spend 3--5\,\% of the time in bright outbursts;
that their most probable flux level 
is F(2--10\,keV)$\sim 1$--$3\times10^{-11}$ erg cm$^{-2}$ s$^{-1}$ 
(unabsorbed, $L\sim 10^{33}$--10$^{34}$ erg s$^{-1}$) 
that their duty cycle of inactivity \cite{Romano2011:sfxts_paperVI} 
is in the range 19--55\,\%, so that these sources accrete matter for most of 
the time \cite{Romano2011:sfxts_paperVI}. 
Fig.~\ref{fig1} and~\ref{fig2}  illustrate our long-term monitoring for the two SXFT prototypes, 
XTE~J1739$-$302 and IGR~J17544$-$2619 (up to $\sim$ MJD 55150) 
and how we keep observing bright outbursts also after the end of the campaign.


Further monitoring programs aimed at SFXTs with known orbital period. 
The first campaign was on IGR~J18483$-$0311 \citep{Romano2010:sfxts_18483}
($P_{\rm orb}\sim 18.5$\,d) which we followed for a whole orbital period, thus allowing us to 
constrain the different mechanisms proposed to explain the SFXT nature. In particular, 
we applied the clumpy wind model for blue supergiants 
\citep{Ducci2009} to the observed X-ray light curve. 
By assuming an eccentricity of $e = 0.4$, 
we could explain the X-ray emission in terms of the accretion from a spherically 
symmetric clumpy wind, composed of clumps with different masses, 
ranging from $10^{18}$ to $\times10^{21}$\,g.  We also monitored 
IGR~J16418$-$4532 ($P_{\rm orb}\sim 3.7$\,d) for three orbital cycles \cite[][]{Romano2012:sfxts_16418} 
and could account for the observed emission with accretion from a spherically symmetric clumpy wind 
(see, also \cite[][]{Esposito12:gamma12}). 

Among the most recent outbursts are those of XTE~J1739$-$302 and IGR~J17544$-$2619 
that triggered the {\it Swift}/BAT on 2011 February 22 and March 24, respectively 
(Fig.~\ref{fig1}, last peaks at $\sim$ MJD 55600, \cite[][]{Farinelli2012:sfxts_paperVIII}) 
and those of IGR~J08408$-$4503 \cite[][]{Mangano12:gamma12}. 


In \cite[][]{Farinelli2012:sfxts_paperVIII}, we find that the spectra of both sources can be
well fitted either with a two-blackbody model, or with a single
unsaturated Comptonization model.
In the latter case, the model can be either a classical static Comptonization model,
such as \comptt, or the recently developed \compmag\ model, which includes thermal
and bulk Comptonization for cylindrical accretion onto a magnetized neutron star.

\begin{figure}
   \includegraphics[angle=270,width=.33\textheight]{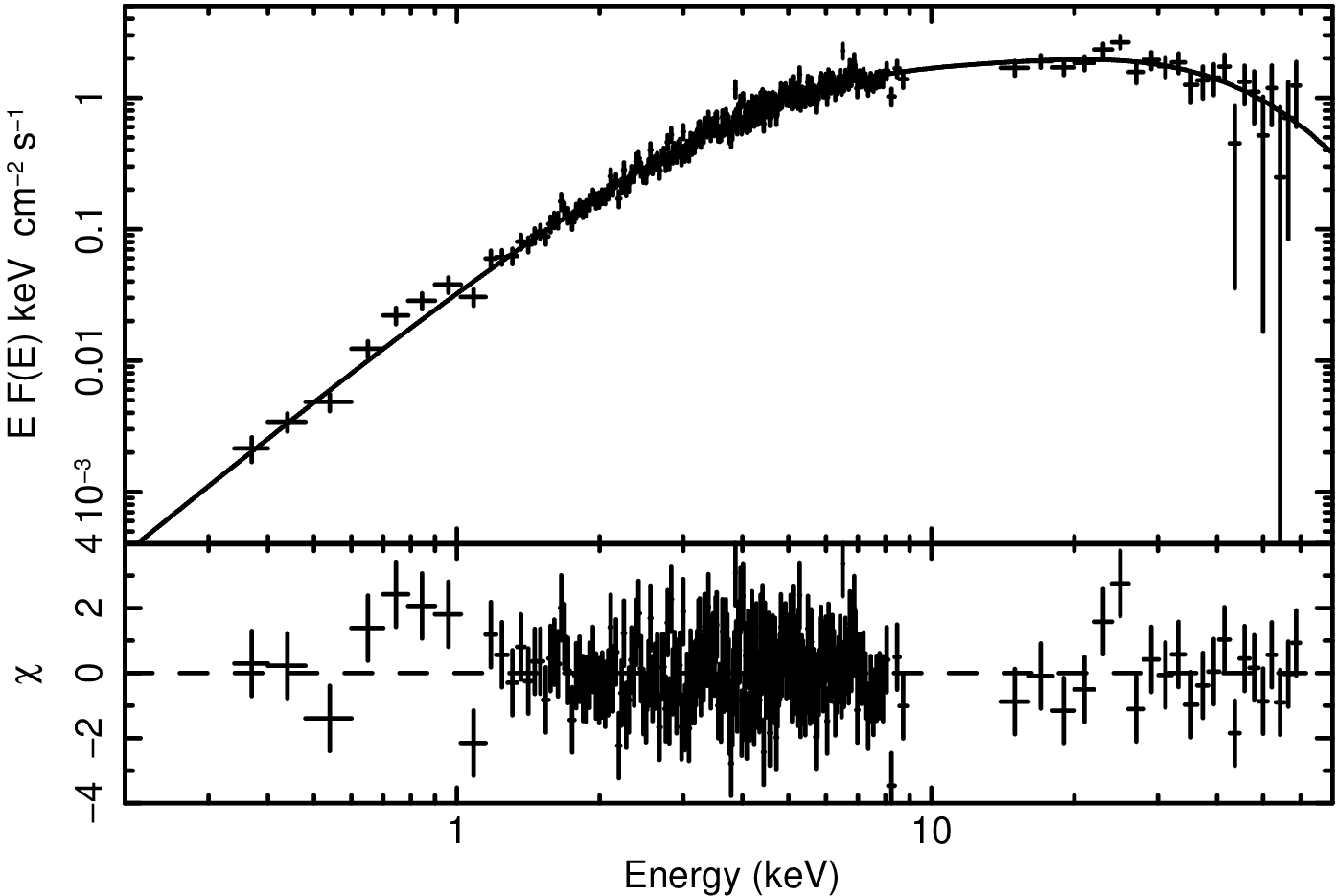}
  \includegraphics[angle=270,width=.33\textheight]{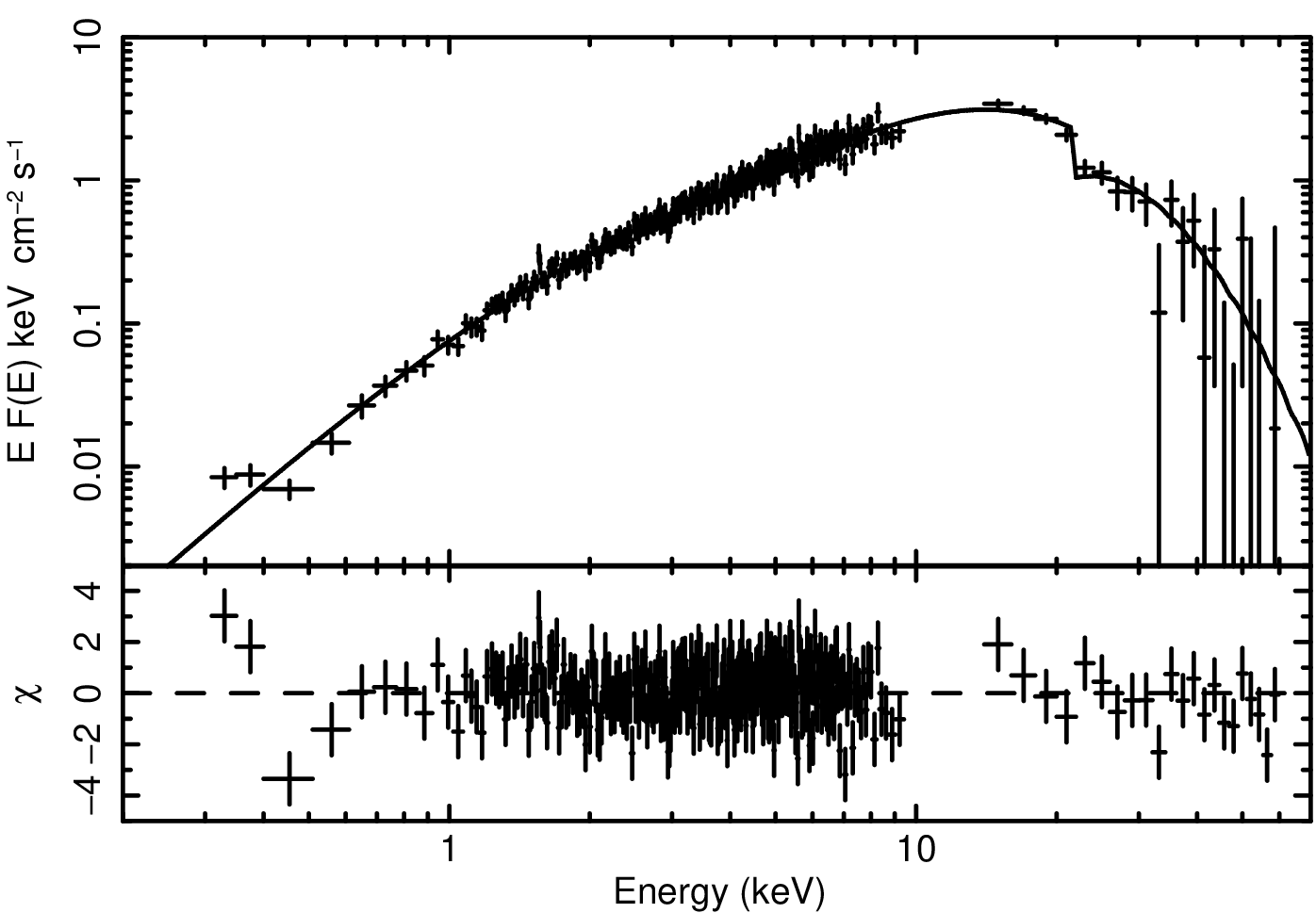}
  \caption{
{\it Left panel:} Absorption-corrected unfolded EF(E) models and spectra, and residuals between the data and
the \compmag\ model in units of $\sigma$ for XTE~J1739$-$302. 
{\it Right panel:}  Absorption-corrected unfolded EF(E) models and spectra, and residuals between the data and
the \compmag\ model in units of $\sigma$ for IGR~J17544$-$2619. 
Adapted from \cite[][]{Farinelli2012:sfxts_paperVIII}. 
}
  \label{fig3}
\end{figure}

We refer to \cite[][]{Farinelli2012:compmag} for a detailed description of the algorithm.  
Here we briefly remind the reader that \compmag\ is based on the solution of the radiative transfer equation
for the case of cylindrical accretion onto the polar cap of a magnetized NS. 
The velocity field of the accreting matter can be increasing towards the NS surface, or it 
may be described by an approximate decelerating profile. 
In the former case, the free parameters are the terminal velocity at the NS surface, $\b0$, and
the index of the law $\beta(z) \propto z^{-\eta}$, while in the second case the law
is given by $\beta(\tau) \propto -\tau$.
The other free parameters of the model are the temperature of the black-body seed photons, $\ktbb$,
the electron temperature and vertical optical depth of the Comptonization plasma $\kte$ and $\tau$, 
respectively, and the radius of the accretion column $r_{\rm 0}$, in
units of the NS \scw\ radius.
The different combinations of these parameters determine the steepness of the spectrum
at high energies and the rollover energy position.
Fig.~\ref{fig3} shows the unfolded EF(E) models and spectra for the \compmag\ model. 


\begin{theacknowledgments}
 We acknowledge financial contribution from the contracts ASI-INAF I/009/10/0 and I/004/11/0.
\end{theacknowledgments}





\IfFileExists{\jobname.bbl}{}
 {\typeout{}
  \typeout{******************************************}
  \typeout{** Please run "bibtex \jobname" to optain}
  \typeout{** the bibliography and then re-run LaTeX}
  \typeout{** twice to fix the references!}
  \typeout{******************************************}
  \typeout{}
 }

\end{document}